\documentclass[twocolumn,aps,prb,superscriptaddress]{revtex4}
\usepackage{graphicx}
\usepackage{amsmath,amsbsy,mathbbol}
\usepackage{psfrag}

\begin{document}
\title{Dynamics of Phononic Dissipation at the Atomic Scale:\\
	Dependence on Internal Degrees of Freedom}

\author{H. Sevin\c{c}li}
\email{sevincli@fen.bilkent.edu.tr}
\affiliation{Department of Physics, Bilkent University, 06800 Ankara, Turkey.}
\affiliation{UNAM-National Nanotechnology Research Center, Bilkent University, 06800 Ankara, Turkey.}
\author{S. Mukhopadhyay}
\affiliation{Department of Physics, Bilkent University, 06800 Ankara, Turkey.}
\affiliation{On leave from Shadan Institute of P.G. Studies, Hyderabad-4, India. }
\author{R. T. Senger}
\affiliation{Department of Physics, Bilkent University, 06800 Ankara, Turkey.}
\affiliation{UNAM-National Nanotechnology Research Center, Bilkent University, 06800 Ankara, Turkey.}
\author{S. Ciraci}
\email{ciraci@fen.bilkent.edu.tr}
\affiliation{Department of Physics, Bilkent University, 06800 Ankara, Turkey.}
\affiliation{UNAM-National Nanotechnology Research Center, Bilkent University, 06800 Ankara, Turkey.}

%\date{\today}

\begin{abstract}
Dynamics of dissipation of a local phonon distribution to the substrate is a key issue in friction between sliding surfaces as well as in boundary lubrication. 
We consider a model system consisting of an excited nano-particle which is weakly coupled with a substrate.
Using three different methods we solve the dynamics of energy dissipation for different types of coupling between the nano-particle and the substrate, where different types of dimensionality and phonon densities of states were also considered for the substrate.
In this paper, we present our analysis of transient properties of energy dissipation via phonon discharge in the microscopic level towards the substrate.
Our theoretical analysis can be extended to treat realistic lubricant molecules or asperities, and also substrates with more complex densities of states.
We found that the decay rate of the nano-particle phonons increases as the square of the interaction constant in the harmonic approximation.
\end{abstract}

\newcommand{\kk}{{\bf k}}
\newcommand{\qq}{{\bf q}}
\newcommand{\ahat}{\hat a}
\newcommand{\bhat}{\hat b}
\newcommand{\aint}{\tilde a}
\newcommand{\bint}{\tilde b}
\newcommand{\ee}{\textrm e}
\newcommand{\real}{\textrm{Re}}
\newcommand{\imag}{\textrm{Im}}

\maketitle

\section{Introduction}\label{section:intro}

Friction between two surfaces in relative motion involves many interesting and complex phenomena induced by the long- and short-range forces, such as adhesion, wetting, atom exchange, bond breaking and bond formation, elastic and plastic deformation\cite{persson_tosatti, bhushan, tomlinson, frenkel_kontriva, bhushan_nature, sutton_pethica, niminen, sorensen, sorensen2, buldum_ciraci, buldum_ciraci_2, buldum_ciraci_3, zhong, tomanek}.
In general, non-equilibrium phonons are generated in   the expense of damped mechanical energy \cite{cieplak, smith, sokoloff1, sokoloff2, sokoloff3, sokoloff4}. 
Dissipation of this excess energy is one of the important issues in dry-sliding friction and lubrication \cite{buldum1, buldum2, buldum3, buldum4}.
Normally, the dissipation of mechanical energy is resulted in heating of parts in relative motion. 
Sometimes, it gives rise to wear and failure due to overheating. 
In general, significant amounts of resources (energy and material) are lost in the course of friction.
One of the prime goals of tribology is to minimize energy dissipation through lubrication. 
Recently, several works have attempted to develop surfaces with superlow friction coefficients. 

The objective of this work is to develop an understanding of phononic energy dissipation during sliding friction, especially to shed some light on the dynamics of discharge of excited phonons of a nano-particle (representing a lubricant molecule or an asperity) deep into the substrate.
This problem has many aspects and the solution will depend on a variety of physical parameters which can be grouped into major categories, such as internal degrees of freedom of the nano-particle, density of substrate phonon modes, the type and strength of coupling between the nano-particle and the substrate, and finally the initial temperatures of the nano-particle and the substrate.
A reliable way of studying this problem is to carry out state-of-the-art molecular dynamics simulations by which one can obtain sample specific results.
To explore the general features of the phononic dissipation, however, we propose an Hamiltonian treatment of the problem.
Since the number of physical ingredients determining the dynamics is considerably large, our strategy will be to focus on them separately.

In this work, we present our conclusions concerning the dependence of phononic dissipation on internal degrees of freedom of the nano-partcile and the substrate, namely on the properties of discrete and continuum densities of phonon modes of the nano-particle and the substrate. 
Furthermore, we will consider two types of coupling between the finite and extended systems.
The effect of initial temperatures of the parties, besides from the effect of temperature difference, is another major topic in its own and we leave that discussion to another paper.
In the present paper, we consider the initial temperatures to be zero and limit our attention to the near-equilibrium case in the weak coupling regime.
Strong coupling regime and non-equilibrium cases will also be treated separately.

The organization of the paper is as follows. 
In Section \ref{section:model}, we describe the physical model. 
The theoretical methods to be used are presented in Section \ref{section:methods}. 
These are the equation of motion (EoM) technique which involves Laplace transforms for the solution of the coupled differential equations for phonon operators, the Fano-Anderson (FA) method which is useful for diagonalizing quadratic Hamiltonians, and Green's Function (GF) method by which we can incorporate the effect of multi-modes into the study. 
The applications of the given methods to different types of coupling and substrates having different densities of states will be presented and discussed in Section \ref{section:applications}. 
These are mainly the increase of decay rate with square of the coupling constant \cite{persson_volokitin}, the interaction-specific dependence of decay rate on the nano-particle mode frequencies and the effect of neighboring modes on decay rate of each other.
Results obtained by the above methods will also be discussed in connection to classical molecular dynamics (MD) simulations in Section \ref{sec:md}.
We summarize our conclusions in Section \ref{section:conclusion}.

\section{The Model}\label{section:model}

\begin{figure}
	\begin{center}
		\includegraphics[scale=0.28]{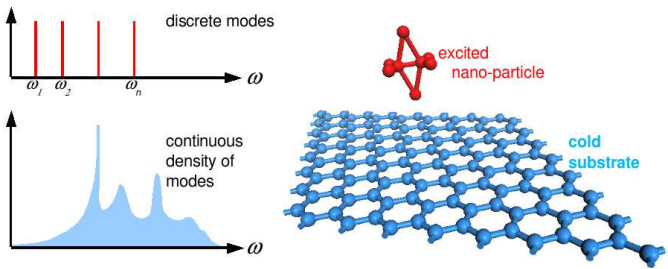}
	\end{center}
	\caption{(Color online) A nano-particle with discrete density of phonon modes is coupled to a substrate having continuous density of modes.}
	\label{fig:iluustration}
\end{figure}

We first consider a nano-particle representing a lubricant molecule or an asperity, which is weakly coupled to a substrate (Fig. \ref{fig:iluustration}). 
The vibrational modes of the nano-particle are excited initially and the excess phonons discharge to the bulk.  
The total Hamiltonian of the system can be written as
\begin{equation}
	H = H_M + H_S + H_{MS}=H_0+H_{MS}
\end{equation}
where $H_M$, $H_S$ are the free phononic Hamiltonians of the nano-particle and the substrate (or bath) respectively. 
We assume that the harmonic approximation is good enough for $H_M$ and $H_S$, and their spectra are known, i.e.
\begin{eqnarray}
	H_M &=& \sum\limits_j \hbar\omega_j\, a_j^+ a_j,\label{eqn:non-overlapping}\\
	H_S &=& \sum\limits_{\bf k} \hbar\omega_{\bf k} b_{\bf  k}^+ b_{\bf k}.
\end{eqnarray}
Here, $\omega_j$ are the frequencies of the nano-particle modes with $a_j$, $a_j^+$ being the corresponding annihilation, creation operators; $\omega_{\bf k}$ are frequencies of the bath modes of wavevectors $\bf k$, and $b_{\bf k}$ , $b_{\bf k}^+$ are the corresponding phonon annihilation, creation operators. 
We have omitted the constant terms as they do not contribute to the dynamics of the system.
We consider a single phonon branch without loosing generality, the formalism can be extended to include multiple branches.
The interaction Hamiltonian $H_{MS}$ is also assumed to be quadratic in phonon operators,
\begin{equation}
	H_{MS}=\sum\limits_{{\bf k},j} \hbar \left( W_{\kk j}b_\kk^+ a_j + h.c.\right) 
\end{equation}
with $W_{\kk j}$ being the coupling coefficient which is a function of $\omega_\kk$ and $\omega_j$, and has the dimesion of angular frequency.

Here we consider two types of coupling. 
The first one is Lorentzian coupling in which the coupling coefficient $W_{\kk j}$ is a Lorentzian with its peak located at $\omega_j$ and has width $\Gamma_j$. 
As long as the coupling between the nano-particle and the substrate is weak, $W_{\kk j}$ will be a peaked function of $\omega_\kk$ and a separate peak will be present around each $\omega_j$. 
Depending on the strength of the interaction, the sharpness of the peaks and the overlap between the neigboring peaks will differ. 
If the coupling is weak enough we may neglect the overlaps, namely
\begin{equation}
	W_{\kk l}^* W_{\kk j} \rightarrow |W_{\kk l}|^2 \delta_{l,j}.
	\label{eqn:model:single_level}
\end{equation}
We assume that the coupling terms of different modes of the nano-particle do not overlap, and hence we can treat each nano-particle mode separately.

For the second type of coupling, we consider the coupling coefficients which scale inversely as the square root of the product of the frequencies of the coupled modes, i.e.
\begin{equation}
	W_{\kk j}=\alpha\left(\omega_\kk \omega_j\right)^{-1/2}.
	\label{eqn:sec:model:sqrt.coupling}
\end{equation}
The coefficient $\alpha$ stands for the strength of the coupling and will depend on the interaction between the nano-particle and the substrate.

\section{Theoretical Methods}\label{section:methods}

\subsection{Equation of Motion  (EoM) Technique}\label{section:eom}

The time dependent occupancies of the nano-particle modes can be obtained using Heisenberg's equation of motion, namely $\dot A(t)=i[H,A(t)]/\hbar$. The equations of motion for the phonon annihilation operators are
\begin{eqnarray}
	\dot a_l(t) &=& -i\omega_l a_l(t) - i\sum\limits_\kk W_{\kk l}^* b_\kk(t),\\
	\dot b_\kk(t) &=& -i\omega_\kk b_\kk(t)-i\sum\limits_j W_{\kk j} a_j(t),
\end{eqnarray}
that is we have coupled differential equations for each operator. Performing Laplace Transformation to both equations, a pair of coupled algebraic equations is obtained
\begin{eqnarray}
	\overline a_l(s)(s+i\omega_l)=a_l(0)-i\sum\limits_\kk W_{\kk l}^*\overline b_\kk(s),\\
	\overline b_\kk(s)\left(s+i\omega_\kk\right)=b_\kk(0)-i\sum\limits_j W_{\kk j}\overline a_j(s),
\end{eqnarray} 
where $s$ is the Laplace frequency.
Solving for $\overline a_l(s)$, one obtains
\begin{eqnarray}
	\overline a_l(s) &=& 
	\frac{a_l(0)}{s+i\omega_l}
	-i\sum\limits_\kk\frac{W_{\kk l}^* b_\kk(0)}
		{(s+i\omega_\kk)(s+i\omega_l)}\nonumber\\
	&-& \sum\limits_{\kk j}\frac{W_{\kk l}^* W_{\kk j} \overline a_j(s)}
		{(s+i\omega_\kk)(s+i\omega_l)}.
\end{eqnarray}

Considering the couplings to be non-overlapping (cf. Eq. (\ref{eqn:model:single_level})) we are left with the relation
\begin{eqnarray}
	\overline a_l(s) =
	\frac{a_l(0)}
	{s+i\omega_l+\sum\limits_\kk\frac{|W_{\kk l}|^2}{s+i\omega_\kk}}
	-i\frac{\sum\limits_\kk\frac{W_{\kk l}^* b_\kk(0)}{s+i\omega_\kk}}
	{s+i\omega_l+\sum\limits_\kk\frac{|W_{\kk l}|^2}{s+i\omega_\kk}}
		\label{eom:a_inverse}
\end{eqnarray}
Having obtained $\overline a_l(s)$ in terms of $a_l(0)$ and $b_\kk(0)$, the inverse Laplace transform will yield $a_l(t)$, thus we can obtain the time dependent occupancy of the $l^{th}$ mode.

We convert the summations into integrals over the substrate modes, and denote them as 
\begin{eqnarray}
	I_l(s)&=&
	\sum\limits_\kk \frac{|W_{\kk l}|^2}{s+i\omega_\kk}=
	\int \frac{d\omega_\kk\, g(\omega_\kk)\,|W_{\kk l}|^2}{s+i\omega_\kk}\\
	J_l(s)&=&
	\sum\limits_\kk\frac{W_{\kk l}^* b_\kk(0)}{s+i\omega_\kk}=
	\int \frac{d\omega_\kk\, g(\omega_\kk)\,W_{\kk l}^* b_\kk(0)}
		{s+i\omega_\kk}
\end{eqnarray}
where $g(\omega_\kk)$ is the phonon density of states for the substrate, $I_l$ and $J_l$ depend on $s$, and $J_l$ is an operator.
We can write Eq. (\ref{eom:a_inverse}) as
\begin{equation}
	\overline a_l(s)=\frac{a_l(0)}{s+i\omega_l+I_l(s)}
		-i\frac{J_l(s)}{s+i\omega_l+I_l(s)}
\end{equation}

The inverse transform of $\overline a_l(s)$ is
\begin{equation}\label{eqn:eom:a_compact}
	a_l(t)=\frac{a_l(0)}{2\pi i}\oint\limits_B \frac{e^{st} ds}{s+i\omega_l+I_l(s)}
		-\frac{1}{2\pi}\oint\limits_B \frac{e^{st} J_l(s) ds}{s+i\omega_l+I_l(s)}
\end{equation}
where the integrals are to be evaluated along the Bromwich contour.	
		
The first and second terms in Eq. (\ref{eqn:eom:a_compact}) stand for contributions  from the initial excitation of the nano-particle and the initial temperature of the  substrate respectively. 
The second term does not contribute to the time dependent occupations of the nano-particle mode, since the initial temperature of the substrate is considered to be zero.

\begin{figure}[h]
	\psfrag{xlabel}[][]{{\bf Time} $(\times 10^{-11}\,sec)$}
	\psfrag{ylabel}[][]{{\bf Occupation} $\langle n_j(t)\rangle$}
        \psfrag{0}[][]{$0$}\psfrag{0.2}[][]{$0.2$}
        \psfrag{0.4}[][]{$0.4$}\psfrag{0.6}[][]{$0.6$}
        \psfrag{0.8}[][]{$0.8$}\psfrag{1}[][]{$1$}
        \psfrag{x 10}[][]{ }
	\begin{center}
		\includegraphics[scale=0.65]{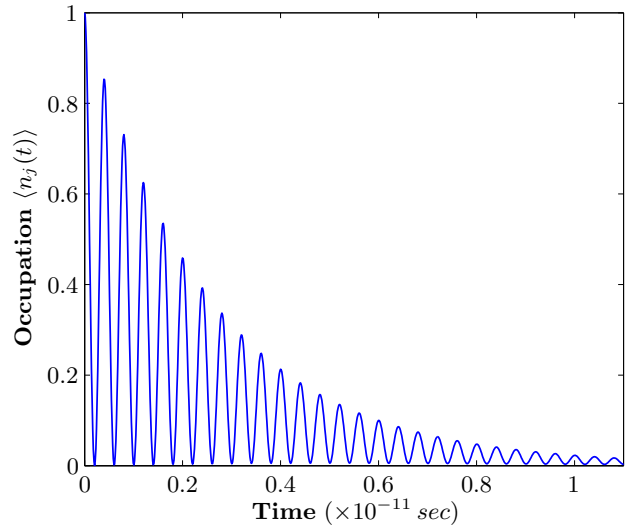}
	\end{center}
	\caption{Decay of a single nano-particle mode $j$ coupled to a 2D-Debye substrate. The coupling is Lorentzian. Occupation $\langle n_j(t)\rangle$ at time $t$ is given relative to the initial occupation $\langle n_j(0)\rangle$.} 
	\label{fig:eom}
\end{figure}

\subsubsection{One- and Two-Dimensional Debye Substrates}

For a specific case, we consider the nano-particle to have a single mode coupled to a one- or two-dimensional Debye substrate which is initially at zero temperature. Assuming Lorentzian coupling, namely
\begin{equation}
	W_{\kk j}^2=\frac{\alpha^2\Gamma}{2\pi}
	\frac{1}{(\omega_\kk-\omega_j)^2+\Gamma^2/4},
	\label{eqn:lorentzian.coupling}
\end{equation}
we have for $I_j(s)$
\begin{eqnarray}
	I_j=\frac{\alpha^2\Gamma c_D}{2\pi i}
	\int\limits_0^{\omega_D} \frac{d\omega_\kk\,\omega_\kk^{d-1}}
	{(\omega_\kk-is)(\omega_\kk-\omega_j-i\frac{\Gamma}{2})
	(\omega_\kk-\omega_j+i\frac{\Gamma}{2})},
\end{eqnarray}
where $d$ is the dimension of the substrate, $c_D$ is the corresponding Debye constant for DOS, $g_d(\omega)=c_D\omega^{d-1}$. $\omega_j$, $\Gamma$ are real and positive, and by definition of Laplace transformation Re$(s)>0$. 
In the weak coupling regime the width of the Lorentzian will be much smaller than $\omega_j$ and $\omega_D$, \cite{lorentzian_aciklama} so we can approximate the above integral by extending the limits of integration to 
$(-\infty,\infty)$ in which case the integral can be evaluated analytically on the complex $\omega_\kk$-plane with the result 
$I_j(s)=\alpha^2c_D \left(\omega_j-i\Gamma/2\right)^{d-1}/
\left(s+i\omega_j+\Gamma/2\right)$. 
Performing the inverse transformation, one finds
\begin{eqnarray}
	\langle n_j(t) \rangle=&& \langle n_j(0) \rangle
	\frac{{\textrm e}^{-\Gamma t/2}}
	{4|\Delta|^2}\nonumber\\
	&\times&
	\left|
	\left(\Gamma+\Delta\right)\ee^{\Delta t/2}-
	\left(\Gamma-\Delta\right)\ee^{-\Delta t/2}
	\right|^2
\end{eqnarray}
where $\Delta^2=\Gamma^2/4-4\alpha^2c_D(\omega_j-i\Gamma/2)^{d-1}$.
%and $\nu_{1,2}=-\Gamma/4\pm\sqrt{\Delta}$.

Domain of applicability of the EoM technique is quite limited due to the fact that inverse Laplace transformation is not always possible neither analytically nor numerically, its advantage is that in certain cases it enables us to get analytical results within some approximations.
Figure \ref{fig:eom} shows the decay of a single nano-particle mode to a 2D-Debye substrate.

\subsection{Fano-Anderson  (FA) Method}\label{section:dressed}

Since the Hamiltonian is quadratic in operators, its solution is equivalent to diagonalizing a matrix. 
Exact diagonalization of such quadratic Hamiltonians was shown to be possible by Fano\cite{fano} and Anderson \cite{anderson} independently and the procedure is widely used in atomic physics, solid state physics, quantum optics, etc. 
Here, we will apply their method to the problem of phononic dissipation. 

Diagonalization of the Hamiltonian is equivalent to finding the dressed operators $\alpha(\omega_\qq)$ such that
\begin{equation}
	H = \sum\limits_\qq \hbar\omega_{\qq}\, \alpha^+(\omega_\qq) \alpha(\omega_\qq).
\end{equation}
Since the bare phonon operators $a_j$, $b_\kk$ form a complete set of operators for the combined system, the dressed operators $\alpha(\omega_\qq)$ can be expanded in terms of the bare operators as
\begin{equation}
	\alpha(\omega_\qq)=
	\sum\limits_j \mu(\omega_\qq,\omega_j)a_j
	+\sum\limits_\kk \nu(\omega_\qq,\omega_\kk)b_\kk
		\label{eqn:dressed:expansion}
\end{equation}
and they satisfy the eigenoperator equation,
\begin{equation}
	\left[ \alpha(\omega_\qq), H \right] = \hbar\omega_\qq\, \alpha (\omega_\qq).
		\label{eqn:dressed:eigen_commutator}
\end{equation}
Conversely, we find the bare operators by the  following expressions in terms of the dressed operators as
\begin{eqnarray}
	a_j &=& \sum\limits_\qq \mu^*(\omega_\qq,\omega_j) \alpha(\omega_\qq),
		\label{dressed:inverse_expansion:a}\\
	b_\kk &=& \sum\limits_\qq \nu^*(\omega_\qq,\omega_\kk)  \alpha(\omega_\qq).
		\label{dressed:inverse_expansion:b}
\end{eqnarray}

Substituting Eq. (\ref{eqn:dressed:expansion}) into Eq. (\ref{eqn:dressed:eigen_commutator}), one ends up with a pair of equations
\begin{eqnarray}
	\mu(\omega_\qq,\omega_j)(\omega_\qq-\omega_j) &=& 
	\sum\limits_\kk \nu(\omega_\qq,\omega_\kk) W_{\kk j}
		\label{dressed:pair:mu}\\
	\nu(\omega_\qq,\omega_\kk)(\omega_\qq-\omega_\kk)&=&
	\sum\limits_j \mu(\omega_\qq,\omega_j)W_{\kk j}^*
		\label{dressed:pair:eta}
\end{eqnarray}
which can be solved self-consistently to obtain $\omega_\qq$. Using Eq. (\ref{dressed:pair:eta}), $\nu$ can be expressed in terms of $\mu$ as
\begin{eqnarray}
	\nu(\omega_\qq,\omega_\kk)&=&
	\left(
		\frac{\mathbb{P}}{\omega_\qq-\omega_\kk}+
		\delta(\omega_\qq-\omega_\kk) z(\omega_\qq)
	\right) \nonumber\\
	&& \times\sum_j \mu(\omega_\qq,\omega_j)W_{\kk j}^*
		\label{dressed:eta:principal}
\end{eqnarray}
where $\mathbb P$ stands for the principal part, and the $\delta$-function term accounts for the contribution from the singularity. Inserting Eq. (\ref{dressed:eta:principal}) into Eq. (\ref{dressed:pair:mu}), the following relation for $\mu(\omega_\qq,\omega_j)$ and $z(\omega_\qq)$ is obtained.
\begin{widetext}
\begin{eqnarray}
	\mu(\omega_\qq,\omega_j)(\omega_\qq-\omega_j) =
	\sum\limits_{\kk l} \frac{{\mathbb P}}{\omega_\qq-\omega_\kk}
	\mu(\omega_\qq,\omega_l) W_{\kk j}  W_{\kk l}^* 
	+ \sum\limits_{\kk l} \delta(\omega_\qq-\omega_\kk) z(\omega_\qq)
	\mu(\omega_\qq,\omega_l) W_{\kk j}  W_{\kk l}^*
\end{eqnarray}
\end{widetext}
If we consider the nano-particle to have a single mode, the relation between $\mu$ and $z$ can be written in a much simpler form and the dissipation of each mode can be treated separately. 
From this point on, we will use a subscript $j$ where necessary denoting that we are working on the dynamics of the $j^{th}$ mode of the nano-particle.

Relying on the above reasoning, $z_j(\omega_\qq)$ can be expressed as
\begin{equation}
	z_j(\omega_\qq)=\frac{\omega_\qq-\omega_j-\sigma_j(\omega_\qq)}
	{g(\omega_\qq)|W_{\qq j}|^2}
		\label{dressed:z}
\end{equation}
$\sigma_j(\omega_\qq)$ being the shift in the $j^{th}$ nano-particle mode
\begin{equation}
	\sigma_j(\omega_\qq) = {\mathbb P}\int 
	\frac	{d\omega_\kk \, g(\omega_\kk) |W_{\kk j}|^2}
		{\omega_\qq-\omega_\kk}.
		\label{dressed:sigma}
\end{equation}
In order to obtain the expansion coefficients $\mu$, $\nu$ phononic commutation relation for $\alpha(\omega_\qq)$ is employed.
\begin{equation}
	[\alpha(\omega_\qq),\alpha^+(\omega_{\qq'})]=\delta_{\qq,\qq'}=
	\frac{\delta(\omega_\qq-\omega_{\qq'})}{g(\omega_\qq)}
		\label{dressed:phononic_commutation}
\end{equation}
Using the expansion in terms of bare operators (\ref{eqn:dressed:expansion}), and Poincare's theorem, i.e.
\begin{eqnarray}
	\frac{\mathbb P}{\omega_\qq-\omega_\kk}\cdot\frac{\mathbb P}{\omega_{\qq'}-\omega_\kk} &=&
	\frac{\mathbb P}{\omega_{\qq'}-\omega_\qq}
	\left(
	\frac{\mathbb P}{\omega_\qq-\omega_\kk}-\frac{\mathbb P}{\omega_{\qq'}-\omega_\kk}
	\right) \nonumber \\
	&+& \pi^2 \delta(\omega_\qq-\omega_\kk)\delta(\omega_{\qq'}-\omega_\kk),
	\label{dressed:poincare}
\end{eqnarray}
the modulus square of $\mu(\omega_\qq,\omega_j)$ is found as
\begin{equation}
	|\mu(\omega_\qq,\omega_j)|^2=\frac{|W_{\qq j}|^2}
	{(\omega_\qq-\omega_j-\sigma_j(\omega_\qq))^2+\pi^2g^2(\omega_\qq)|W_{\qq j}|^4}
		\label{dressed:mu_kare}
\end{equation}

Since the Hamiltonian is diagonal with annihilation, creation operators $\alpha(\omega_\qq)$, $\alpha^+(\omega_\qq)$ and eigenfrequencies $\omega_\qq$, the time dependence of the dressed annihilation operator is
\begin{equation}
        \alpha(\omega_\qq,t)=
		\mu(\omega_\qq,\omega_j)a_j e^{-i\omega_\qq t}
		+\sum\limits_\kk \nu(\omega_\qq,\omega_\kk)b_\kk e^{-i\omega_\qq t}
        \label{eqn:dressed:expansion_time}
\end{equation}
Correspondingly, the time dependence of the nano-particle annihilation operator reads (cf. Eq. (\ref{dressed:inverse_expansion:a}))
\begin{eqnarray}
	a_j(t)=\int d\omega_\qq \, g(\omega_\qq) \, \mu^*(\omega_\qq,\omega_j)\, \alpha(\omega_\qq)\, e^{-i\omega_\qq t}.
\end{eqnarray}

Hence the occupancy of the $j^{th}$ mode is
\begin{widetext}
\begin{eqnarray}
	\langle n_j(t) \rangle = \langle  n_j(0) \rangle
		\left|
			\int d\omega_\qq \, g(\omega_\qq) \,
			|\mu(\omega_\qq,\omega_j)|^2  e^{-i\omega_\qq t}
		\right|^2
	+\int d\omega_\kk g(\omega_\kk) \langle  n_\kk(0) \rangle 
		 \left|
			\int d\omega_\qq \, g(\omega_\qq) \,
			\mu^*(\omega_\qq,\omega_j)\,\nu^*(\omega_\qq,\omega_\kk)
			e^{-i\omega_\qq t}
		 \right|^2 
\end{eqnarray}
\end{widetext}

Due to the finite range of substrate DOS $g(\omega_\kk)$, the integrals involved in the FA Method are bounded, the method allows us to perform calculations for any $g(\omega_\kk)$ and any type of coupling for a single nano-particle mode. 
In the FA method, the time dependent occupation is again separable as contributions from the initial temperature of the nano-particle and that of the substrate.

\subsection{Green's Function (GF) Method}\label{section:green}

The FA Method is applicable for any coupling type and any density of states for the substrate as long as we consider a single nano-particle mode. 
For the multi-mode case, the effect of neighboring modes cannot be resolved within the above procedures, so we need to develop a more generalized method by which we can consider such effects.
For this purpose we employ GFs. 
Initially the substrate temperature is zero and the phonon modes of the nano-particle are empty except for the excitations which do not necessarily obey Bose-Einstein distribution.
That is the initial occupation of a nano-particle mode is not a function of temperature.
Therefore we make use of zero temperature GFs instead of Matsubara formalism,
\begin{eqnarray}
       d(j,t-t')&=&-i\langle T_t \ahat_j(t)\ahat_j^+(t')\rangle \\
       D(\kk,t-t')&=&-i\langle T_t \bhat_\kk(t)\bhat_\kk^+(t')\rangle
\end{eqnarray}
where $T_t$ is the time-ordering operator, and the operators in Heisenberg picture are distinguished by a hat.

Since each term in the interaction Hamiltonian includes odd number of nano-particle operators, only the even order terms contribute in the expansion.
The first contribution due to the interaction is the second order term,
\begin{widetext}
\begin{eqnarray}
	\sum\limits_{\kk}W_{j\kk}^2 \int\limits_{-\infty}^{\infty} dt_1
	\int\limits_{-\infty}^{\infty} dt_2 \,
	d^{(0)}(j,t-t_1)D^{(0)}(\kk,t_1-t_2)d^{(0)}(j,t_2-t')
\end{eqnarray}
\end{widetext}
whose Fourier Transform gives
\begin{equation}
	d^{(0)}(\omega_j)^2\sum\limits_\kk W_{j\kk}^2 D^{(0)}(\omega_\kk)=
	d^{(0)}(\omega_j)^2 \Sigma^{(2)}(\omega_j)
\end{equation}
with $\Sigma^{(2)}(\omega_j)$ being the second order self-energy.

\subsubsection{Nano-Particle with Single Mode}

First, we need to relate the GF and FA methods, therefore we consider a single nano-particle mode, in which the higher order terms can be expressed in terms of the second order self-energy and the free GF as
\begin{figure}
	\psfrag{(a)}[][]{\bf (a)}
	\psfrag{(b)}[][]{\bf (b)}
	\psfrag{t}[][]{$t$}
	\psfrag{t1}[][]{$t_1$}
	\psfrag{t2}[][]{$t_2$}
	\psfrag{t3}[][]{$t_3$}
	\psfrag{t4}[][]{$t_4$}
	\psfrag{tn-2}[ ][ ]{$t_{2n-2}$ }
	\psfrag{tn-1}[][]{$t_{2n-1}$}
	\psfrag{tn}[][]{$t_{2n}$}
	\psfrag{t'}[][]{$t'$}
	\psfrag{j}[][]{$j$}
	\psfrag{j1}[ ][ ]{$j_1$}
	\psfrag{jn-1}[ ][ ]{$j_{n-1}$}
	\psfrag{k1}[][]{$\kk_1$}
	\psfrag{k2}[][]{$\kk_2$}
	\psfrag{kn}[][]{$\kk_n$}
        \includegraphics[scale=0.425]{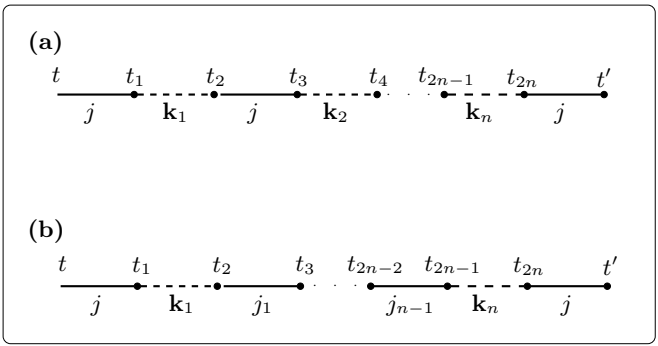}
        \caption{Diagrams of order $2n$. Solid lines are the phonon lines of the nano-particle where the dashed lines are that of the substrate. (a) Diagram for the case of single nano-particle mode, $j$. $\kk_i$ stand for the substrate modes.
	(b) Diagram of order $2n$ when there exists multiple modes ($j_i $) for the nano-particle.}
        \label{fig:feynman}
\end{figure}
\begin{eqnarray}
	d(\omega_j)=d^{(0)}(\omega_j)
	\left(
		1+d^{(0)}\Sigma^{(2)}+(d^{(0)}\Sigma^{(2)})^2+...
	\right).
\end{eqnarray}
For weak coupling, the above series can be written as
\begin{eqnarray}
	d(\omega_j,\omega_\qq)&=& 
	\frac{d^{(0)}(\omega_j,\omega_\qq)}
	{1-d^{(0)}(\omega_j,\omega_\qq)\Sigma^{(2)}(\omega_j,\omega_\qq)},
\end{eqnarray}
hence, the retarded function becomes
\begin{eqnarray}
        d^R(\omega_j,\omega_\qq) &=& 
	\frac{1}{\omega_\qq-\omega_j-\Sigma^{(2)}(\omega_j,\omega_\qq)}.
\end{eqnarray}

The real and imaginary parts of the second order self-energy can be separated 
\begin{eqnarray}
	\Sigma^{(2)}(\omega_j,\omega_\qq)=
	{\mathbb P}\int \frac{d\omega_\kk\,g(\omega_\kk)W_{j\kk}^2}
	{\omega_\qq-\omega_\kk}
	-i\pi g(\omega_\qq)W_{j\qq}^2
	\label{eqn:green_real.imag_self2}
\end{eqnarray}
where $g(\omega)$ stands for substrate density of states and $\mathbb P$ is for principal part of the integral, and the spectral function is obtained
\begin{equation}
        A(j,\omega_\qq)=
        \frac{-2\, \textrm{Im}\,\Sigma^{(2)}(j,\omega_\qq)}
	{\left(\omega_\qq-\omega_j-\textrm{Re}\,\Sigma^{(2)}
	(j,\omega_\qq)\right)^2+    
	\left(\textrm{Im}\,\Sigma^{(2)}(j,\omega_\qq)\right)^2}.
	\label{eqn:green_spectral}
\end{equation}
We should note here the correspondence between the FA  and the GF methods. 
The real part of the  second order self-energy $\Sigma^{(2)}$ is equal to the shift in the $j^{th}$ mode of the nano-particle, $\sigma_j$, obtained previously using the FA method, and the square of the imaginary part of $\Sigma^{(2)}$ is $(\pi g(\omega_\qq)W_{\qq j})^2$. 
That is, the FA expansion coefficient $\mu$ finds its expression in terms of the spectral function as
\begin{equation}
	|\mu(\omega_\qq,\omega_j)|^2=
	\frac{A(j,\omega_\qq)}{2\pi g(\omega_\qq)}.
\end{equation}
The time dependent GF can be written in terms of the spectral function and the time dependent occupancy of the $j^{th}$ mode is obtained as
\begin{equation}
        \langle n_j(t) \rangle= \langle n_j(0) \rangle
	\left|
        \int \frac{d\omega_\qq}{2\pi}\, A(j,\omega_\qq)\,
        \textrm e^{-i\omega_\qq t}
	\right|^2.
\end{equation}

\subsubsection{Nano-Particle Having Multi-Modes}

As long as the Hamiltonian is quadratic, the primitive vertex, from which all diagrams are to be constructed, will consist of two phonon lines. That is, each interaction point is the intersection of two phonon lines. Since the interaction Hamiltonian relates a nano-particle mode to a substrate mode only, each vertex contains one nano-particle phonon line and a substrate phonon line. So the diagram of any order can be constructed (see Figure \ref{fig:feynman}.b). 
Having obtained the diagrammatic expansion for any order $2n$, under certain conditions about the coupling type, the $(2n)^{th}$ order self-energy term can be expressed in terms of the second order term and the free GF of the nano-particle modes. 
If the fraction $W_{\kk_1 j_1}/W_{\kk_1 j_2}$ is independent of $\kk_1$, the exact form of the self-energy for the multi-mode case reads (see Appendix \ref{section:appendix}),
\begin{eqnarray}
	\Sigma(j,\omega_\qq)=
	\frac{\Sigma^{(2)}(j,\omega_\qq)}
	{1-\frac{\Sigma^{(2)}(j,\omega_\qq)}{W_{j\qq}^2} \sum\limits_{j'\neq j}
	W_{j'\qq}^2 d^{(0)}(j',\omega_\qq)}
\end{eqnarray}
Once the self-energy is found, the spectral function, therefore the time dependent occupancy of the nano-particle modes can be calculated.

\section{Discussions and Numerical Results}\label{section:applications}

In this section, we will apply the above methods to Lorentzian and square-root coupling of nano-particles to substrates having 1D and 2D Debye DOS and discuss the dependence of decay rate on properties of coupling, and the internal parameters of the nano-particle and the substrate.
The lowest vibrational frequency of a nano-particle increases as its size decreases.
In the weak coupling regime, the width of the spectral function $A(j,\omega_\qq)$ will be small compared to $\omega_\qq$ provided that $\omega_j$ is not close to zero, which is already satisfied for nano-particles.
In this limit, the imaginary part of the self-energy can be interpreted as twice the decay rate, $\gamma_j=\imag\,\Sigma(j,\omega_\qq)/2$.
Therefore, the dependence of decay rate on the interaction type and strength as well as on the frequency of the nano-particle can be obtained from the spectral function.

For the case of Lorentzian coupling, the interaction strength $\alpha$ is linear with $W_{\kk j}$ which shows that the decay rate increases with $\alpha^2$. If the coupling is a function of the separation between interacting atoms of the substrate and the nano-particle only, the coupling has the form of Eq. (\ref{eqn:sec:model:sqrt.coupling}) with $\alpha$ being proportional to a spring constant $k_{int}$ connecting the interacting atoms.
Since the spectral function scales with $\alpha^2$, decay rate increases with $k_{int}^2$ for inverse-square-root coupling case.
The $k_{int}^2$ law was previously obtained using elastic continuum model for phononic dissipation at physisorption systems \cite{persson_volokitin}.

The dependence on the nano-particle mode frequency is a key issue we wish to emphasize in phononic energy dissipation. In Lorentzian coupling case, the decay rate is determined by the width of the Lorentzian rather than the frequency. On the other hand, for inverse-square-root coupling (cf. Eq. (\ref{eqn:sec:model:sqrt.coupling})), it is inversely proportional to the nano-particle mode frequency $\omega_j$. 
\begin{figure*}
	\includegraphics[scale=0.7]{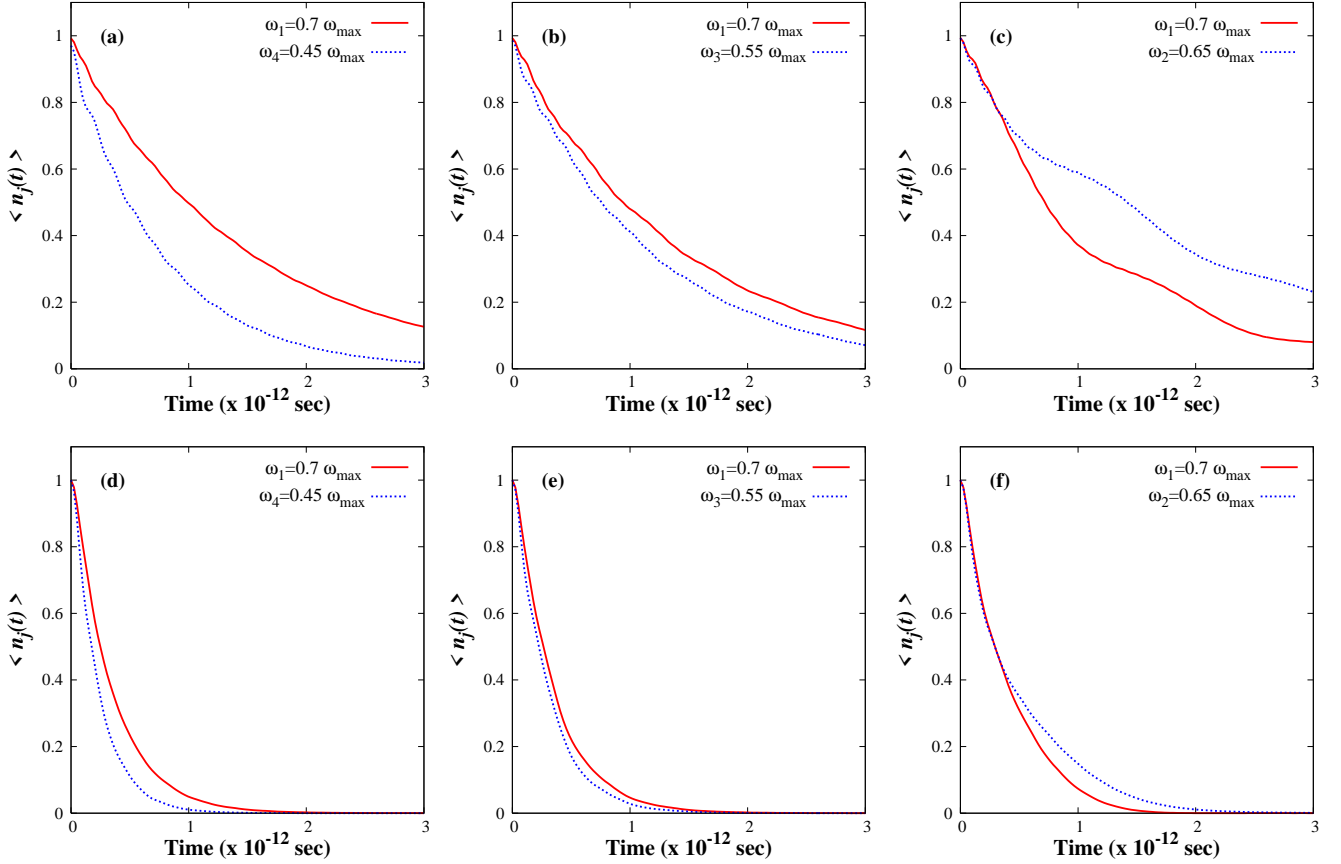}
	\caption{(Color online) Effect of neighboring modes. Figures (a), (b), (c) are for 1D-Debye DOS, and (d), (e), (f) are for 2D-Debye DOS with nano-particle vibration frequencies $\omega_1= 0.7\,\omega_{max}$, $\omega_2= 0.65\,\omega_{max}$, $\omega_3= 0.55\, \omega_{max}$, $\omega_4= 0.45 \,\omega_{max}$.
Figures (a) and (d), (b) and (e) and (c) and (f) show dissipation of phonon occupation for the pairs $(\omega_1,\omega_4)$, $(\omega_1,\omega_3)$, and  $(\omega_1,\omega_2)$ respectively.}
	\label{fig:2modes}
\end{figure*}

\begin{figure}
	\includegraphics[scale=0.85]{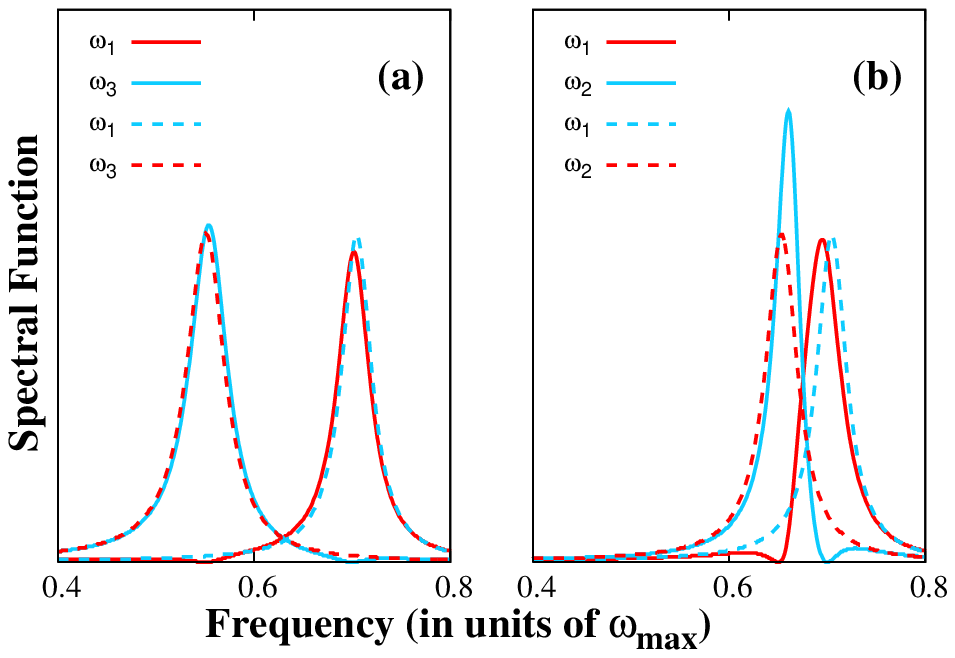}
	\caption{(Color online) Effect of a neighboring mode on the spectral function. Dashed curves are the single mode spectral functions whereas the solid curves are spectral functions in the existence of a neighboring mode. (a) Spectra of $\omega_1=0.7\,\omega_{max}$ and $\omega_3=0.55\,\omega_{max}$ for both cases are almost the same. (b) Spectra of $\omega_1=0.7\,\omega_{max}$ and $\omega_2=0.65\,\omega_{max}$ get narrowed and distorted when single mode condition is relaxed.}
	\label{fig:narrow}
\end{figure}

It is evident from equations (\ref{dressed:mu_kare}) and (\ref{eqn:green_spectral}) that phonons in mode $\omega_j$ decay faster as the substrate DOS at the center of the peak, $\omega_j-\real\,\Sigma(j,\omega_\qq)$, increases. 
A crucial consequence of dependence on substrate DOS is that, if the DOS at the peak of the spectral function tends to zero, the spectral function (and $|\mu(\omega_\qq,\omega_j)|^2$) has the form of a $\delta$-function. 
In the language of dressed modes, this corresponds to a localized mode, i.e. it does not decay at all. Such localized states are also known to occur in e.g. solid state physics \cite{anderson} and atomic physics \cite{fano}.
For weak coupling, the real part of the self-energy is small, so the peak of the spectral function is not altered significantly from its original position $\omega_j$.
In other words, lying outside the continuum of substrate modes, it is unlikely to be shifted into the range where it can decay and vice versa.
Another important effect about DOS dependence takes place when the spectral peak coincides with a van-Hove singularity of the substrate DOS, by which the decay rate is enhanced abruptly.

We investigate the effect of a neighboring mode within square-root coupling in 1D Debye and 2D Debye substrate densities of states using GF method.
We consider four nano-particle modes, $\omega_1=0.7\,\omega_D$, $\omega_2=0.    65\,\omega_D$, $\omega_3=0.55\,\omega_D$ and $\omega_4=0.45\,\omega_D$. 
The effect is analyzed pairwise, namely, we consider $(\omega_1,     \omega_2)$, $(\omega_1, \omega_3)$ and $(\omega_1, \omega_4)$ as the   nano-particle modes keeping other parameters unchanged. 
That is, we keep $\omega_1$ constant while changing the second mode and   investigate dependence of decay of $\omega_1$-mode on the separation from the second nano-particle mode. 
For both 1D-Debye (Figure \ref{fig:2modes} (a) (b) and (c)) and 2D-Debye (Figure \ref{fig:2modes} (d), (e), and (f)) cases we observe that the decay of excited modes gain a retardation as the mode frequencies get closer. 
A second  behavior is the enhancement of fluctuations during decay as the mode frequencies get closer.
Both behaviors can be understood in terms of the spectral functions. 
In Figure (\ref{fig:narrow}.a) spectral functions of $\omega_1-$ and $\omega_3-$modes are plotted for single-mode (dashed curves) and multi-mode (solid curves) GF calculations. 
It is seen that the overlap is negligible and the spectra are not changed considerably.
When the modes are closer (Figure \ref{fig:narrow}.b) the single-mode spectra (dashed curves) have finite overlap, correspondingly the multi-mode spectral functions effect each other. 
The Lorentzian shape is distorted and the peak of $\omega_2-$mode is enhanced.
These result in retardation and fluctuations during decay.
More precisely, the finite overlap of spectra lets the nano-particle to gain phonons back which are previously discharged to the substrate.
This phonon exchange process continues during the dissipation and gives rise to retardations and fluctuations observed in Figure (\ref{fig:2modes}).

\subsection{Molecular Dynamics Simulations}\label{sec:md}

In order to provide another basis on which the results obtained from the quantum treatment of the problem can be discussed, we refer to a classical treatment of the problem.
We use a simple but effective approach, where we consider the nano-particle/substrate as a cluster/lattice of masses and harmonic springs in the first nearest neighbor approximation, and with the lattice having different dimensionalities.
The interaction is described by a harmonic spring between an atom of the  nano-particle and a substrate atom.
Using the dynamical matrix, the eigenmodes of the isolated nano-particle are determined and initial energy is loaded to desired modes by giving the initial velocities to the atoms in correspondence with the modes.
In the presence of the interaction between the nano-particle and the substrate, the differential equations and hence the motion of all atoms are determined in discrete time steps which are on the order of a femtosecond.

Since the above stated classical version of the problem is not an exact analog to the quantum one, and due to quantum vs. classical natures of the two, we compare and contrast the basic features of the results emerging from them.
To be specific, we expect to verify qualitative issues like dependence on interaction strength, nano-particle mode frequency and the effect of neighboring modes.

In agreement with the earlier prediction based on the elastic continuum model \cite{persson_volokitin}, and the result previously obtained using the spectral function, the dependence of decay rate on the interaction strength obeys $k_{int}^2$ law for weak coupling.
Likewise, the dependence on internal degrees of freedom of the nano-particle verifies the previously obtained result. 
\begin{figure}
	\begin{center}
        \includegraphics[scale=0.85]{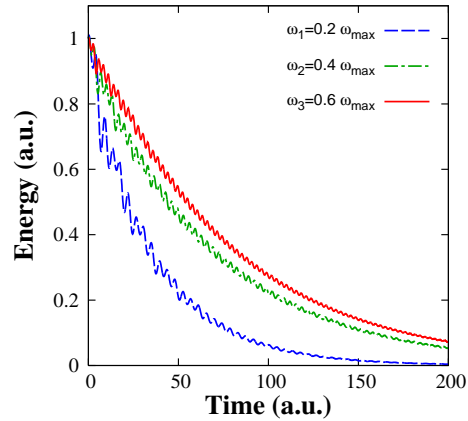}
        \caption{(Color online) MD simulation: 1D-substrate and different nano-particle modes. Higher frequency modes decay slower.}
	\label{fig:md_kmol_1d}
	\end{center}
\end{figure}
Keeping the coupling strength constant, higher frequency modes decay slower(Figure \ref{fig:md_kmol_1d}).
We should note that the substrate DOS in the neighborhood of the nano-particle mode frequency also affects the decay rates.
Using a 1D substrate and choosing the nano-particle modes away from the maximum frequency of the substrate, the effect of substrate DOS is minimized. 
Although the density of substrate phonon modes is higher for higher frequencies, the decay rate decreases due to nano-particle mode frequency dependence.

Another property of the dissipation process becomes apparent when the dynamics is analyzed when the nano-particle has one and two vibrational modes.
We consider a diatomic molecule and a linear triatomic molecule, which have one and two vibrational modes along the molecular axis, where the interaction is also along the molecular axis.
The effect of neighboring mode can be be analyzed by setting one of the modes of the triatomic molecule at the same frequency with the frequency of the diatomic one. 
Exciting only the common frequency of both diatomic and triatomic molecules, we compare the decay rates keeping the interaction and substrate parameters fixed.
In the weak coupling regime, it is observed that the decay rate of the common mode does not change appreciably.
Moreover, exciting both vibrational modes of the triatomic molecule does not effect the decay rate of the common mode to a great extend. 
This property becomes more apparent as the coupling strength is weakened.
Since the vibrational modes of a triatomic molecule along the molecular axis are well-separated this result is expected in the light of GF solution of the quantum Hamiltonian.
The mode localization effect is also tested using classical MD simulations. 
A molecular mode whose frequency lies above the maximum frequency of the substrate has a small but yet finite decay rate.

\section{Conclusion}\label{section:conclusion}

A theoretical understanding of the phononic dissipation from a nano-structure weakly coupled to a substrate has been developed using three different methods. 
The EoM technique is able to yield analytical results, but has a limited range of applicability because of the fact that inverse Laplace transformation is not always possible. 
On the other hand, FA diagonalization is possible for any type of substrate density of states and any type of coupling, but is restricted to single nano-particle mode considerations only. 
Using GFs, the effect of neighboring nano-particle modes can also be investigated. 
It is shown that stronger the coupling, faster is the rate of dissipation. 
Since the width of the spectrum of a single nano-particle mode scales with the value of the substrate DOS at the shifted frequency of the nano-particle mode, we observe that a single nano-particle mode coupled to a 2D-Debye substrate decays faster than the one coupled to a 1D-Debye substrate. 
This situation can be reversed for those frequencies for which 1D-DOS is higher than the 2D-DOS, namely for low frequencies (larger nano-particles). 
That is, at frequencies at which 1D-DOS has higher values than 2D-DOS, decay rate of a mode coupled to the 1D substrate will be higher than that of the mode coupled to 2D substrate provided that the remaining factors are kept identical. 
Presence of neighboring nano-particle modes effect each other's decay rate when their spectral functions have an appreciable overlap. 
Transitions between nano-particle modes take place via the substrate modes, therefore retardation as well as fluctuations become important when the modes are close enough.

\begin{acknowledgments}
This work was supported by The Scientific and Technological Research Council of Turkey through Grant No. TBAG-104T537.
R. T. S. acknowledges financial support from T\"UBA-GEB\.IP.
\end{acknowledgments}

\appendix

\section{Self-Energy For Multi-Modes}\label{section:appendix}
When $W_{\kk,j_1}/W_{\kk,j_2 }$ is independent of $\kk$, the fourth and sixth order contributions can be written as
\begin{eqnarray}
	\Sigma^{(4)}(j,\omega_q)=
	\frac{1}{W_{\qq j}^2}
	\left( \sum\limits_{j_1} d^{(0)}(j_1,\omega_q)W_{\qq j_1}^2\right)
	\left( \Sigma^{(2)}(j,\omega_\qq)\right)^2\\
	\Sigma^{(6)}(j,\omega_q)=
	\frac{1}{W_{\qq j}^4}
	\left( \sum\limits_{j_1} d^{(0)}(j_1,\omega_q)W_{\qq j_1}^2\right)^2
	\left( \Sigma^{(2)}(j,\omega_\qq)\right)^3
\end{eqnarray}
By mathematical induction, the $(2n)^{th}$ term is found as
\begin{eqnarray}
	\Sigma^{(2n)}(j,\omega_q)&=&
        \frac{1}{W_{\qq j}^{2(n-1)}}
        \left( \sum\limits_{j_1} d^{(0)}(j_1,\omega_q)W_{\qq j_1}^2\right)^{n-1}\nonumber\\
	&&\times
        \left( \Sigma^{(2)}(j,\omega_\qq)\right)^n.
\end{eqnarray}


\begin{thebibliography}{99}
		
% intro
\bibitem{persson_tosatti}
\emph{Physics of Sliding Friction}, Vol. 311 of 
\emph{NATO Advanced Studies Institute, Series E, Applied Sciences},
edited by B. N. J. Persson and E. Tosatti
(Kluwer, Dordrecht, 1996)

\bibitem{bhushan}
\emph{Micro/Nanotribology and its Applications}, Vol. 330 of
\emph{NATO Advanced Studies Institute, Series E, Applied Sciences},
edited by B. Bhuhan
(Kluwer, Dordrecht, 1997)

\bibitem{tomlinson}
G. A. Tomlinson, Philos. Mag. \textbf{7}, 905 (1929) 

\bibitem{frenkel_kontriva}
J. Frenkel and T. Kontorova, Phys. Z. Sowjetunion \textbf{13}, 1 (1938).

\bibitem{bhushan_nature}
B. Bhushan, J. N. Israelachvili, and U. Landman, Nature (London) \textbf{347}, 607 (1995).

\bibitem{sutton_pethica}
A. P. Sutton and J. B. Pethica, J. Phys.: Condens. Matter \textbf{2}, 5317 (1990).

\bibitem{niminen}
J. A. Nieminen, A. P. Sutton, and J. B. Pethica, Acta Metall. Mater. \textbf{40}, 2503 (1992).

\bibitem{sorensen}
M. R. Sorensen, K. W. Jacobsen, and P. Stoltze, Phys. Rev. B \textbf{53}, 2101 (1996)

\bibitem{sorensen2}
M. R. Sorensen, K. W. Jacobsen, and H. Jonsson, Phys. Rev. Lett. \textbf{77}, 5067 (1996).

\bibitem{buldum_ciraci}
A. Buldum and S. Ciraci, Phys. Rev. B \textbf{55}, 2606 (1997).

\bibitem{buldum_ciraci_2}
A. Buldum and S. Ciraci, Phys. Rev. B \textbf{55}, 12892 (1997). 

\bibitem{buldum_ciraci_3}
A. Buldum, S. Ciraci, and I.P. Batra, Phys. Rev. B \textbf{57}, 2468 (1998).

\bibitem{zhong}
W. Zhong and D. Tomanek, Phys. Rev. Lett. \textbf{64}, 3054 (1990).

\bibitem{tomanek}
D. Tomanek, W. Zhong, and H. Thomas, Europhys. Lett. \textbf{15}, 887 (1991).

\bibitem{cieplak}
M. Cieplak, E. D. Smith, and M. O. Robins, Science \textbf{265}, 1209 (1994).

\bibitem{smith}
E. D. Smith, M. O. Robbins, and M. Cieplak, Phys. Rev. B \textbf{54}, 8252 (1996).

\bibitem{sokoloff1}
J. B. Sokoloff, Phys. Rev. B \textbf{42}, 760 (1990).

\bibitem{sokoloff2}
J. B. Sokoloff, Phys. Rev. B \textbf{42}, 6745(E) (1990).

\bibitem{sokoloff3}
J. B. Sokoloff, Phys. Rev. B \textbf{51}, 15573 (1995).

\bibitem{sokoloff4}
J. B. Sokoloff, Phys. Rev. Lett. \textbf{71}, 3450 (1993).

\bibitem{buldum1}
A. Buldum and S. Ciraci Phys. Rev. B \textbf{60}, 1982 (1999)

\bibitem{buldum2}
A. Buldum, D. M. Leitner, and S. Ciraci Phys. Rev. B \textbf{59}, 16042 (1999)

\bibitem{buldum3}
A. Buldum, S. Ciraci, and I. P. Batra Phys. Rev. B \textbf{57}, 2468 (1998)

\bibitem{buldum4}
A. Buldum and S. Ciraci Phys. Rev. B \textbf{55}, 2606 (1997)


% dressed
\bibitem{fano}
U. Fano, Phys. Rev. \textbf{124}, 1866 (1961).

\bibitem{anderson}
P. W. Anderson, Phys. Rev. \textbf{124}, 41 (1961).

% application
\bibitem{persson_volokitin}
B. N. J. Persson, A. I. Volokitin, 
\emph{Physics of Sliding Friction},  Vol. 311 of
\emph{NATO Advanced Studies Institute, Series E, Applied Sciences},
edited by B. N. J. Persson and E. Tosatti,
(Kluwer, Dordrecht, 1996), pp. 253-264.

\bibitem{lorentzian_aciklama}
As long as $(\omega_j/\Gamma)\gg 1$ and $(\omega_D/\Gamma)\gg 1$ the Lorentzian is localized into the range where the Debye-DOS is finite, so no extra contribution raises with extending the limits of integration.

\end{thebibliography}
\end{document}